\documentclass[sansserif,mathserif]{article}
\usepackage{pgf,tikz}
\usepackage{pgfplots}
\usepackage{amsmath}
\usepackage{rotating}

\usetikzlibrary{calc,shadows}
\usetikzlibrary{pgfplots.groupplots}

\usepgfplotslibrary{groupplots} 
\usetikzlibrary{pgfplots.groupplots}

\usepgfplotslibrary{fillbetween}
\pgfplotsset{compat=1.10}
\usepackage{xspace}

\newcommand\M{M} 
\newcommand\N{N} 
\newcommand\len{L} 

\usepackage{hyperref}
\hypersetup{
    bookmarks=true,         
    unicode=false,          
    pdftoolbar=true,        
    pdfmenubar=true,        
    pdffitwindow=false,     
    pdfstartview={FitH},    
    pdfsubject={Subject},   
    pdfproducer={Producer}, 
    pdfnewwindow=true,      
    colorlinks=true,       
    linkcolor=red,          
    citecolor=green,        
    filecolor=magenta,      
    urlcolor=cyan           
}

\usepackage[
backend=bibtex, 
url=false,isbn=false,doi=false,
date=year,
hyperref=auto,
style=alphabetic,
natbib=true,
sorting=nty,
firstinits=true,
maxnames=2, maxbibnames=10,
]{biblatex}

\usepackage{hyperref}

\newcommand{\vm}{\mathbf{m}}

\newcommand\figurechannelmodel{
\begin{tikzpicture}[scale=0.7,>=latex]
\draw[gray] (-1,-1.5) rectangle (10.7,2.5);

\begin{scope}[xshift=-3cm]
\node at (-1.2,0) {\small 010110};
\draw[->, out=0,in =-180] [->] (0,0) to node[above,swap] {
\hspace{0.8cm}
\parbox{2cm}{\parbox{1.9cm}{\small encode}}
} (2,0);
\end{scope}

\begin{scope}[xshift=-1.1cm]
\draw[->, out=0,in =-180] [->] (12,0) to node[above,swap] {
\hspace{0.8cm}
\parbox{2cm}{\parbox{1.9cm}{\small decode}}
} (13.8,0);
\node at (14.9,0) {\small 010110};
\end{scope}

\node at (-0,0.4) {\tiny \textcolor{blue}{ACATACGT}};
\node at (-0,0) {\tiny \textcolor{red}{CATGTACA}};
\node at (-0,-0.4) {\tiny \textcolor{brown}{GCTATGCC}};

\draw[->] [->] (1,0) to (1.8,0);
\node at (1.5,0) [above,rotate=60] {\hspace{1.1cm}\scriptsize synthesis};

\begin{scope}[xshift=-10cm,yshift=3cm]
\draw[thick, drop shadow, fill=white] (13,-3) circle (1cm);
\draw[very thick, blue] (13.4,-3) to (13.7,-3);
\draw[very thick, red] (13,-2.45) to (12.8,-2.2);
\draw[very thick, red] (12.2,-3.4) to (12.5,-3.3);
\draw[very thick, brown] (12.7,-3.1) to (13.0,-3);
\draw[very thick, brown] (13,-2.2) to (13.3,-2.3);
\end{scope}

\draw[->] (4.3,0) to node[above,swap,rotate=60] {\hspace{1.5cm}\scriptsize amplification
} (5.1,0);
\begin{scope}[xshift=-6.7cm,yshift=3cm]
\draw[thick, drop shadow, fill=white] (13,-3) circle (1cm);
\draw[very thick, blue] (13.1,-3) to (13.2,-3.3);
\draw[very thick, blue] (13.4,-3) to (13.7,-3);
\draw[very thick, blue] (13.4,-2.7) to (13.5,-2.45);
\draw[very thick, blue] (12.4,-2.7) to (12.5,-2.45);
\draw[very thick, blue] (12.8,-2.7) to (13.1,-2.7);
\draw[very thick, red] (12.8,-3.8) to (13.1,-3.7);
\draw[very thick, red] (12.2,-3.4) to (12.5,-3.3);
\draw[very thick, red] (12.9,-3.4) to (12.65,-3.2);
\draw[very thick, red] (13,-2.45) to (12.8,-2.2);
\draw[very thick, red] (13.5,-3.2) to (13.5,-3.5);
\draw[very thick, brown] (12.5,-3.1) to (12.5,-2.8);
\draw[very thick, brown] (12.7,-3.1) to (13.0,-3);
\draw[very thick, brown] (13.7,-3.2) to (13.9,-3);
\draw[very thick, brown] (13,-2.2) to (13.3,-2.3);
\draw[very thick, brown] (12.4,-3.5) to (12.7,-3.6);
\end{scope}

\draw[->] [->] (7.7,0) to (8.6,0);
\node at (8.3,0.2) [above,rotate=60] {\hspace{1cm}\scriptsize sequencing};
\begin{scope}[xshift=-1.1cm]
\node at (10.7,0.8) {\tiny \textcolor{blue}{ACATACGT}};
\node at (10.7,0.4) {\tiny \textcolor{red}{CATGTACA}};
\node at (10.7,0) {\tiny \textcolor{red}{CATGTACA}};
\node at (10.7,-0.4) {\tiny \textcolor{blue}{ACATACGT}};
\node at (10.7,-0.8) {\tiny \textcolor{blue}{ACATACGT}};
\end{scope}

\begin{scope}[yshift=-4.8cm,xshift=-0.9cm,>=latex]
\node [rotate=90] at (5.8,2.5) {$=$};

\draw[gray] (-2,-2) rectangle (13.7,2);

\node at (1,0.5) {\scriptsize \textcolor{blue}{ACATACGT}};
\node at (1,0) {\scriptsize \textcolor{red}{CATGTACA}};
\node at (1,-0.5) {\scriptsize \textcolor{brown}{GCTATGCC}};

\draw[->] [->] (4,-0.3) to node[above,swap] {sample \& disturb} (8,-0.3);

\draw[decoration={brace,mirror,raise=5pt,amplitude=4pt},decorate]
  (-0.2,-0.5) -- node[below=6pt,yshift=-0.15cm] {$L$} (2.2,-0.5);

\draw [decorate,decoration={brace,amplitude=4pt},xshift=-10pt,yshift=0pt]
(0,-0.7) -- (0,0.7) node [black,midway,xshift=-0.4cm] 
{$M$};

\node at (10.7,1) {\scriptsize \textcolor{blue}{ACATACGT}};
\node at (10.7,0.5) {\scriptsize \textcolor{red}{CATGTACA}};
\node at (10.7,0) {\scriptsize \textcolor{red}{CATGTACA}};
\node at (10.7,-0.5) {\scriptsize \textcolor{blue}{ACATACGT}};
\node at (10.7,-1) {\scriptsize \textcolor{blue}{ACATACGT}};

\draw [decorate,decoration={brace,amplitude=4pt,mirror},xshift=1pt,yshift=0pt]
(12,-1.3) -- (12,1.3) node [black,midway,xshift=0.4cm] 
{$N$};

\end{scope}

\end{tikzpicture}
}

\usepackage{lipsum}

\newcommand\blfootnote[1]{%
  \begingroup
  \renewcommand\thefootnote{}\footnote{#1}%
  \addtocounter{footnote}{-1}%
  \endgroup
}

\title{A Characterization of the DNA Data Storage Channel}
\date{\today}

\begin{document}

\begin{center}

{\bf{\LARGE{
A Characterization of the DNA Data Storage Channel
}}}

\vspace*{.2in}

{\large{
Reinhard Heckel$^{\ast}$, Gediminas Mikutis$^{\dagger}$, and Robert N. Grass$^{\dagger}$
\blfootnote{$^\ast$To whom correspondence should be addressed (\href{rh43@rice.edu}{rh43@rice.edu})}
}}

\vspace*{.05in}

\begin{tabular}{c}
Department of Electrical and Computer Engineering, Rice University$^\ast$
\end{tabular}

\begin{tabular}{c}
Department of Chemistry and Applied Biosciences, ETH Zurich$^\dagger$
\end{tabular}

\vspace*{.1in}

\today

\vspace*{.1in}

\begin{abstract}
Owing to its longevity and enormous information density, DNA, the molecule encoding biological information, has emerged as a promising archival storage medium. 
However, due to technological constraints, data can only be written onto many short DNA molecules that are stored in an unordered way, and can only be read by sampling from this DNA pool. Moreover, imperfections in writing (synthesis), reading (sequencing), storage, and handling of the DNA, in particular amplification via PCR, lead to a loss of DNA molecules and induce errors within the molecules.
In order to design DNA storage systems, a qualitative and quantitative understanding of the errors and the loss of molecules is crucial.
In this paper, we characterize those error probabilities by analyzing data from our own experiments as well as from experiments of two different groups. We find that errors within molecules are mainly due to synthesis and sequencing, while imperfections in handling and storage lead to a significant loss of sequences.
The aim of our study is to help guide the design of future DNA data storage systems by providing a quantitative and qualitative understanding of the DNA data storage channel.
\end{abstract}

\end{center}

\section{Introduction}


Recent years have seen an explosion in the amount, variety, and importance of data created, and much of that data needs to be archived. As an example, CERN, the European particle research organization has spent billions of Dollars to generate more than 100 petabytes of physical data which it archives for analysis by future generations of scientists. However, standard storage media such as optical discs, hard drives, and magnetic tapes only guarantee data lifetimes of a few years. This has spurred significant interest in new storage technologies. Fueled by the excitement about its longevity and enormous information density, Deoxyribonucleic acid (DNA), a molecule that caries the genetic instruction of living organisms, has emerged as a promising archival storage medium. 

At least since the 60s, computer scientists and engineers have dreamed of harnessing DNA's storage capabilities~\cite{neiman_fundamental_1964,baum_building_1995}, but the field has only been developed in recent years: In 2012 and 2013 groups lead by Church~\cite{church_next-generation_2012} and Goldman~\cite{goldman_towards_2013} stored about a megabyte of data in DNA and in 2015 Grass et al.~\cite{grass_robust_2015} demonstrated that millenia long storage times are possible by information theoretically and physically protecting the data. Later, in the same year, Yazdi et al~\cite{yazdi_rewritable_2015} showed how to selectively access files, and in 2017, Erlich and Zielinski~\cite{erlich_dna_2016} demonstrated that DNA achieves very high information densities. In 2018, Organick et al.~\cite{organick_scaling_2017} scaled up those techniques and successfully stored and retrieved more than 200 megabytes of data.

DNA is a long molecule made up of four nucleotides (Adenine, Cytosine, Guanine, and Thymine) and, for storage purposes, can be viewed as a string over a four-letter alphabet.
However, there are several technological constraints for writing (synthesizing), storing, and reading (sequencing) DNA.
The perhaps most significant one is that in practice it is difficult to synthesize strands of DNA significantly longer than one-two hundred nucleotides. While there are approaches that generate significantly longer strands of DNA, those are based on writing short strands of DNA and stitching them together~\cite{gibson_creation_2010}, which is currently not a scalable approach. 
As a consequence, all recently proposed systems~\cite{church_next-generation_2012, goldman_towards_2013,grass_robust_2015,bornholt_dna-based_2016,erlich_dna_2016,yazdi_rewritable_2015,organick_scaling_2017} 
stored information on DNA molecules of one-two hundred nucleotides. 
The second technological constraint is that the DNA molecules are stored in a pool and cannot be spatially ordered.
We do not have random access to the DNA fragments in the pool, and can therefore not choose which DNA fragments to read. 
Accessing the information is done via state-of-the-art sequencing technologies (including Illumina and third-generation sequencing technologies such as nanopore sequencing).
This corresponds to (randomly) sampling and reading molecules from the DNA pool. 
In practice, sequencing is preceded by potentially several cycles of Polymerase Chain Reaction (PCR) amplification. In each cycle, each DNA molecule is replicated by a factor of about two, but that number depends on the PCR method and varies by sequence. 
Thus, the proportions of molecules in the pool depend on the synthesis method, the PCR steps, and the decay of DNA during storage. 
In summary, in the process of synthesizing, storing, handling, and sequencing, the following errors occur:
\begin{enumerate}
\item[i.] Molecules might not be successfully synthesized, and some might be synthesized many more times than others. 
Current synthesis technologies generate not only one but several thousand to millions copies of a strand, which can all contain possibly different errors. 
\item[ii.] During storage, DNA decays, which results in a loss of molecules.
\item[iii.] Since reading amounts to drawing from the pool of molecules, we only see a fraction of the molecules that are in the pool. This fraction depends on the distribution of the molecules in the pool and the number of draws (i.e., reads) we take. 
\item[iv.] Synthesizing and sequencing DNA may lead to insertions, deletions, and substitutions of nucleotides in individual DNA molecules.
\end{enumerate}

\begin{figure}[h!]
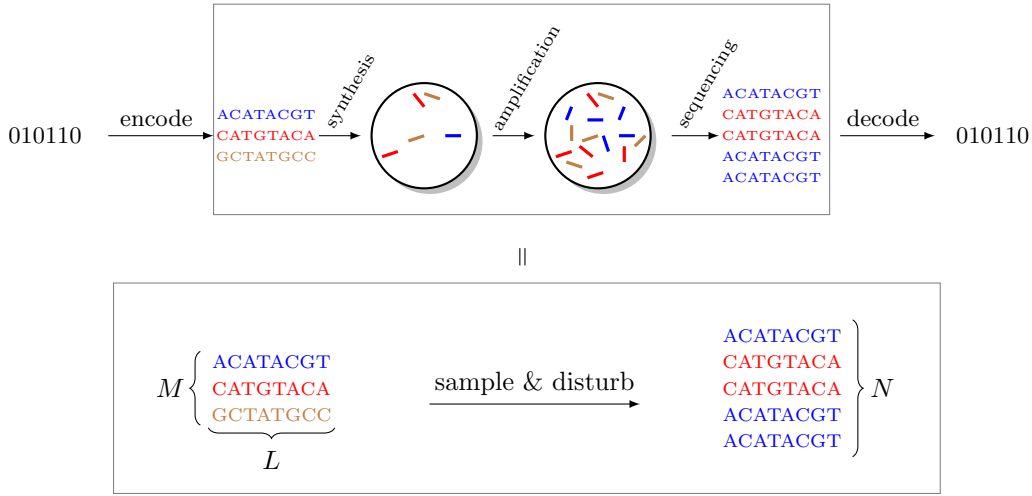

\begin{center}
\figurechannelmodel
\end{center}
\caption{
\label{fig:channelmodel}
Channel model for DNA storage systems. Only short molecules can be synthesized, and of each molecule a large number of copies is generated in the synthesis process. 
For reading, the data is first amplified and then sequenced.
Abstractly, the input to the channel is a multi-set of $\M$ length-$\len$ DNA molecules, while the output is a multi-set of $\N$ draws from the pool of DNA molecules that is disturbed by insertions, substitutions, and deletions. 
The sampling distribution as well as the process inducing errors in individual molecules account for errors in synthesis, storage, and sequencing.
}
\vspace{-2mm}
\end{figure}

Given these constraints, a good abstraction of the DNA data storage channel that captures the essential parts of a DNA storage system is to view the input as a multiset of $M$ DNA molecules of length $L$, and the output as sampling $N$ times independently from the multiset, and then disturbing the sampled molecules with insertions, deletions, and substitutions, to account for errors within individual molecules. 
See Figure~\ref{fig:channelmodel} for an illustration. 
The sampling distribution and distribution of insertions, deletions, and substitutions determine the channel and thus how much data can be stored. 
A statistical understanding of those distributions and the processes that lead to those distributions is crucial for designing good DNA data storage systems, and in particular for designing good encoder/decoder pairs. Providing such an understanding is the goal of this paper.

Note that, for a given channel---determined by the handling procedures, experimental designs, synthesis and sequencing tools, and storage time---the goal of an encoder/decoder pair is to maximize the number of bits per nucleotide, $ML$, while guaranteeing successful decoding.
The goal when designing a DNA data storage system is to minimize the cost for storing data, or to maximize the number of bits per \emph{total number of nucleotides} stored, which is typically significantly larger than the number of nucleotides in the multiset, $ML$. 
Therefore, it is important to understand how different parameters of the channel, such as storage times and physical density (i.e., number of copies of the DNA molecules in the pool) affect the error distributions. 
To determine the best trade-off for maximizing density or minimizing costs, an understanding of the error statistics and how they change with different experimental setups is important. 

The aim of this paper is to obtain a both quantitative and qualitative understanding of the DNA data storage channel, which amounts to quantifying the sampling distribution as well as the distribution of errors within the molecules, for different experimental setups, and assign, when possible, error sources to processes such as reading or writing the DNA. 
As currently the cost of well designed experiments to analyze all those errors in a fully controlled manner is very high, we chose to use data for quantifying error statistics from the DNA data storage literature, as well as complement that data with data from some of our own experiments. 

We start by giving an explanation of the basic chemical processes involved, together with a short theoretical discussion of the potential errors. Next, we identify differences between the various performed experiments and assign the observed differences in error probabilities to the individual steps. This should not only give theoreticians estimates of the overall expected errors, but also directions on how errors could be experimentally avoided.

\section{\label{sec:errsources}Error sources}

Errors in DNA synthesis, storage, and sequencing can have several origins, which we discuss in this section. 

\subsection{Errors during DNA synthesis}

The currently utilized synthesis strategies work on solid surfaces (chips), on which one end of the DNA molecule is attached, and building blocks (nucleotides) are added one by one by a chemical process~\cite{kosuri_large-scale_2014}. These synthesis methods are currently able to generate up to 1,000,000 (typically distinct) sets of DNA strands per chip. 
Note that current technologies are not able to generate single DNA strands; they generate sets of DNA strands. Each set typically consists of millions copies (femtomoles) \cite{schmidt_scalable_2015}, and is generated on a geometrically limited 2-D surface of the chip (i.e., a spot). 
Nucleotides are directed to these spots by light via optically sensitive protection groups \cite{sack_simultaneous_2013,singh-gasson_maskless_1999}, or are directed electrochemically with electrodes via pH sensitive protection groups \cite{maurer_electrochemically_2006}, 
or are directed with printing technology by using a movable printing head~\cite{leproust_synthesis_2010}. 
During the chemical addition of new nucleotides on a growing DNA strand, various errors can occur: a nucleotide may not be positioned \emph{where it should} (resulting in a deletion), a nucleotide might be positioned \emph{where it should not} (resulting in an insertion), and a nucleotide other than the intended one is added (resulting in a substitution). 
Moreover, the growing strand may be terminated, meaning that the chemical reactivity at the end of the growing strand is lost, and nucleotides can no longer be added to this strand in later steps. The probability of this termination reaction---which is around 0.05\%~\cite{leproust_synthesis_2010}---limits the length of DNA strands that can be generated, since the longer the target sequence, the more sequences at a spot do not reach the target length.
Also note that typically not all individual growing strands at the same spot (corresponding to the same target sequence) undergo the same error, meaning that for every target sequence variations may be present. 

Depending on the chosen synthesis method, the number of DNA copies generated per spot may be unevenly distributed. It can easily be imagined that spots on the chip-edge may have a different synthesis yield (number of complete DNA strands synthesized per sequence), that some DNA sequences intrinsically have higher yields than others (e.g., synthesis limited by self-binding of sequences) and that physical imperfections on the chip surfaces limit the ideal synthesis on some chip locations. 

All chemical synthesis methods generate single stranded DNA (ssDNA), which is chemically detached from the chip surface after the synthesis procedure to generate a DNA pool in aqueous solution.
In order to clean up the synthesis pool, i.e., remove non-complete sequences, and to generate double stranded DNA, polymerase chain reaction (PCR) is performed. During this step utilizing biotechnological enzymes, only DNA sequences which have correct primers on both ends are amplified about 10,000 fold over about 15 cycles (different labs utilize slightly different procedures). This process dilutes out non-complete strands (DNA strands that do not have primers on both ends can not be amplified). 

Although PCR by itself is understood as a high-fidelity process and thus has few errors~\cite{cline_pcr_1996,lubock_systematic_2017}. 
PCR is known to have a preference for some sequences over others, which may further distort the copy number distribution of individual sequences~\cite{ruijter_amplification_2009,pan_dna_2014,warnecke_detection_1997,caldana_quantitative_2007}. 
Thus, each cycle of PCR in expectation multiplies each molecule by a certain number which lies typically slightly below two and is sequence dependent. 

\subsection{Errors during DNA storage\label{sec:storagechemerr}}

During DNA storage, as well as during any DNA processing step, such as removing DNA from the chips, and during heating intervals in PCR, the DNA strands are prone to chemical decay of DNA, especially hydrolytic damage. 
The main effects of hydrolytic damage are depurination~\cite{lindahl_rate_1972}, which eventually results in strands breaking~\cite{suzuki_mechanistic_1994}, and deamination of cytosine (C), in the C-G basepair, resulting in uracil (an U-G basepair)~\cite{lindahl_heat-induced_1974}. 

Using current DNA reading technologies which involve DNA amplification stages, namely standard PCR in sample preparation and bridge amplification during Illumina sequencing, any DNA strand that has undergone strand-breakage following depurination is no longer read.
This is due to the fact that broken strands do not have the required amplification primers on both ends of the DNA strand, therefore do not amplify and as a consequence are diluted out during the amplification stages.

\begin{figure}
\begin{center}
\includegraphics[width=0.85\textwidth]{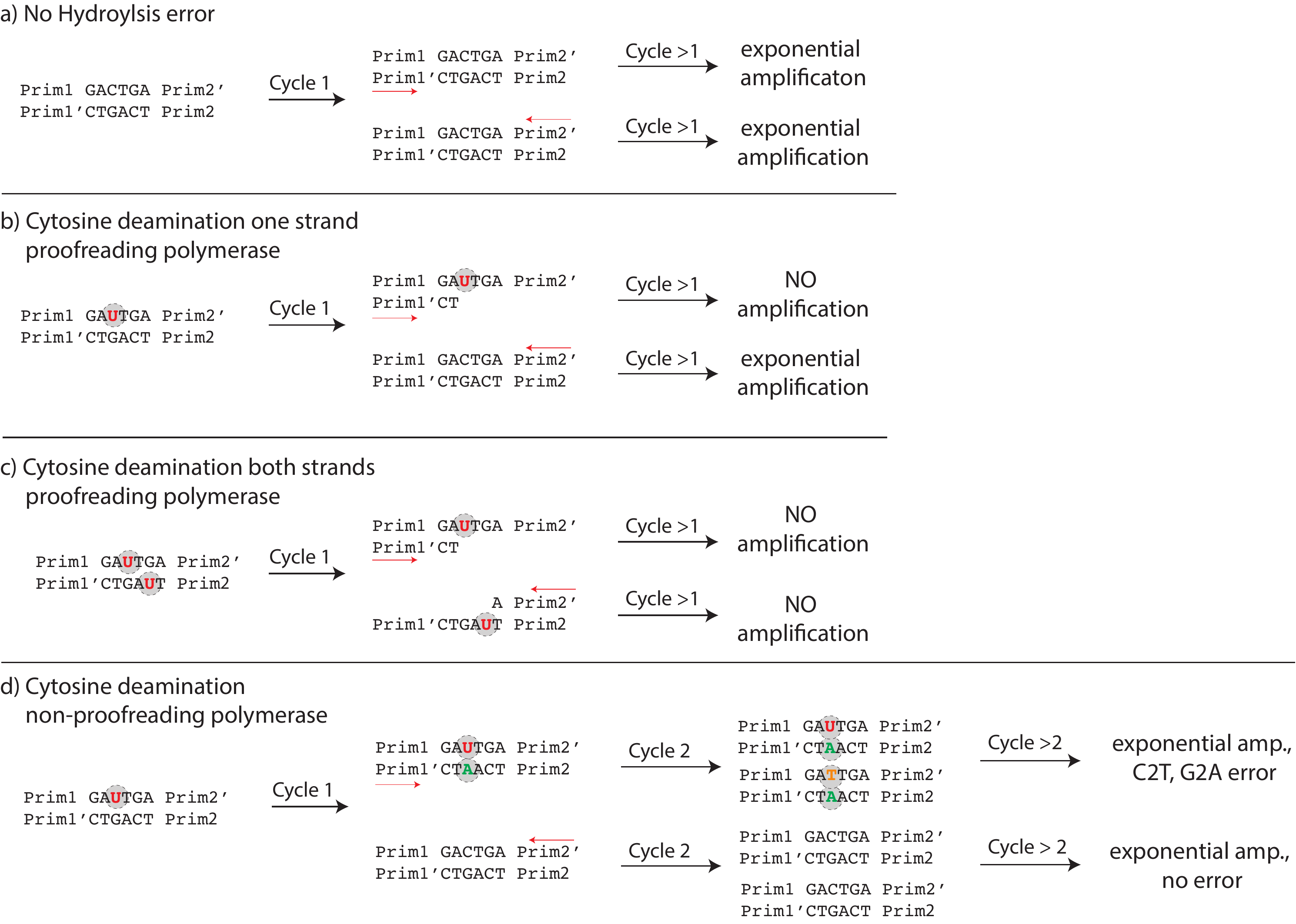}
\end{center}
\caption{\label{fig:proofreading}
Deamination of cytosine results in the transformation of cytosine to uracyl. Most proof-reading PCR enzymes cannot copy past an uracyl. As a consequence any strand containing a de-aminated cytosine is not amplified (b). If both strands of a dsDNA molecule contain uracyl moieties, all information originally contained in the molecule is lost (diluted out) during PCR (c). If non-proof reading enzymes are used for PCR (e.g. taq polymerase), any uracyl is misinterpreted by the enzyme as a thymine, introducing an adenine on the growing complementary strand, resulting in C2T and G2A errors.
} 
\end{figure}

The effect of hydrolytic deamination is less extreme, and depends on the choice of enzymes utilized during subsequent PCR steps. Proof-reading enzymes have a significantly different effect on the errors introduced during deamination (i.e., U-G basepairs) than non-proof-reading enzymes (3' to 5' exonuclease, high-fidelity). For most proof reading enzymes, the amplification reaction stalls at an U nucleotide in the first PCR cycle, and no complete copy of a sequence comprising a U nucleotide is formed. These incomplete DNA copies are then not further amplified in the preceding PCR cycle, due to a lack of primers on both ends of the sequence, and are therefore diluted out. 
On the complementary strand (having a G), the sequence still amplifies. 
In this context, the enzyme removes the errors which have occurred during DNA storage. However, if a DNA strand has at least one deamination (C to U reaction) on both strands, the whole information stored in the sequence is lost (diluted out), as the amplification reaction stalls at the U nucleotide at both strands and neither are amplified (see Figure~\ref{fig:proofreading}). 
If non-proof reading enzymes are utilized,  the U-G basepair is in half amplified to the correct C-G basepair, and in half amplified incorrectly as a T-A basepair, formally resulting in a C2T error.

In summary, storage can lead to significant loss of whole DNA molecules, as well as to substitution errors in the DNA molecules.

\subsection{Errors during DNA sequencing 
\label{sec:errorsseq}}

The errors during DNA sequencing depend on the technology that is used. 
The currently most utilized sequencing platform is Illumina's, and in this paper we only consider datasets that have been sequenced with this technology, although recently DNA data storage readout has also been performed with nanopore sequencing~\cite{organick_scaling_2017,yazdi_portable_2017}. 
Ilumina errors are not random, but are strand specific~\cite{schirmer_illumina_2016}, and the error rates are higher at the end of a read. 
According to a recent study~\cite{schirmer_illumina_2016}, the substitution reading error rates are about 0.0015-0.0004 errors per base. We remark that the exact number, however, depends on the particular dataset. 
Insertions and deletions are significantly less likely, specifically they are on the order of $10^{-6}$. 

Two properties that raise the error probability for most sequencing (and synthesis) technologies are high GC content and long homopolymer stretches (e.g., GGGGGG)
\cite{ross_characterizing_2013,schwartz_accurate_2012}. 
In more detail, the paper \cite[Fig.~5]{ross_characterizing_2013} reports that in particular the substitution and deletion error rates increase significantly for homopolymer stretches longer than six. 
Moreover, molecules with high GC content exhibit high dropout rates and PCR errors are significantly more likely in regions with a high GC content~\cite{schwartz_accurate_2012, erlich_dna_2016}.
For example, the paper \cite[Fig.~3]{ross_characterizing_2013} reports that the coverage obtained by common sequencing technologies such as Illumina's HiSeq technology is significantly lower for fragments with GC content smaller than $20\%$ and larger than about $75\%$. For reasons described in the previous section, any DNA strand that is broken, or for any other reason, does not have correct sequencing primers on both ends of the DNA strand can not be read by Illumina sequencing technologies, as the initial phase of the Illumina sequencing procedure involves bridge-amplification~\cite{bentley_accurate_2008}.



\section{Error statistics}

Here, we estimate the error probabilities of the DNA data storage experiments performed by three different research groups and discuss the allocation of the errors to individual parts (writing, storage, reading) of the DNA storage model. We start with describing the data sets we consider. We proceed with discussing errors on a molecule level (substitutions, insertions, and deletions) as well as the distribution of molecules that we observe on the output of the channel. 

\subsection{\label{sec:data}Data sets}

\newcommand\origDNA{High PR\xspace}
\newcommand\origOld{High PR $4t_{1/2}$\xspace}
\newcommand\buffer{Low PR\xspace}
\newcommand\bufferOld{Low PR $4t_{1/2}$\xspace}
\newcommand\Erlich{Erlich\xspace}
\newcommand\Goldman{Goldman\xspace}

In this paper, we analyze datasets from our group as well as datasets from Goldman's~\cite{goldman_towards_2013} and Erlich's groups~\cite{erlich_dna_2016}. 
Each dataset was obtained by synthesizing DNA molecules, storing the molecules, and sequencing them. For each dataset, we are given the channel input consisting of the DNA molecules to be synthesized, and the channel output, in the form of forward and backward reads from Illunmina sequencing technology.
For each dataset, we obtained DNA molecules by stitching the one-sided reads together using the FLASH \cite{magoc_flash_2011} algorithm. 
We set the maximal overlap of the reads to the target length, which is known in each experiment, the minimal overlap to $87\%$ of the target length, and finally, the maximum allowed ratio between the number of mismatched based pairs and the overlap length to $20\%$. 

The datasets differ in the number of \textbf{strands synthesized}, the \textbf{target length} of each of the sequences, the \textbf{synthesis method}, the expected \textbf{decay} due to thermal treatment (modeling accelerated aging), measured in the percent of the original DNA that remains intact after heat treatment, 
the \textbf{physical redundancy}, which is the expected number of copies of each strand in the pool of DNA, the number of \textbf{PCR cycles}, which is the total number of PCR cycles conducted from synthesis to sequencing, the \textbf{synthesis method}, and finally the \textbf{read coverage} which is number of reads per strand that has been synthesized. 
We briefly summarize how the datasets have been obtained below, and the key parameters in Table~\ref{tab:pardata}. 

\begin{itemize}
\item 
\citet{goldman_towards_2013} synthesized 153,335 DNA molecules, each of length 117, containing no homopolymers. 
DNA was synthesized using the Agilent Oligo Library Synthesis (OLS) Sureprint technology  process, and sequenced with Illumina HighSeq 2000 by taking forward and backward reads of length exactly 104.
\item \citet{erlich_dna_2016} synthesized 72,000 DNA molecules, each of length 152, containing no homopolymers longer than two. 
DNA was synthesized with Twist Bioscience technology, and sequenced with Illumina Miseq V4 techonlogy, taking forward and backward reads of length exactly 151. 
We also consider data from the dilution experiment in~\citet{erlich_dna_2016}, which varies the physical redundancy by factors of ten starting from approximately $10^7$ by diluting the DNA (Erlich D1-7).

\item \citet{grass_robust_2015} synthesized 4991 DNA molecules, each of length 117, containing no homopolymers longer than three. 
DNA was synthesized with Customarray, and the DNA was read after storing it for a negligible amount of time and read again after thermal treatment corresponding to four half-lifes. Both dataset have rather high physical redundancy and are therefore denoted by \origDNA
and \origOld. 
The DNA was sequenced with Illumina Miseq 2x150bp Truseq technology. 

\item In another dataset from our group, we synthesized again 4991 DNA molecules, each of length 117, containing no homopolymers longer than three. In contrast to \origDNA and \origOld, we diluted the DNA so that the physical redundancies are low.
We again sequenced the original DNA as well as DNA after thermal treatment corresponding to four half-lifes; the resulting datasets are denoted by \buffer and \bufferOld. See Appendix~\ref{sec:bufferdescription} for a more detailed description.
\end{itemize}

\begin{table}[h!]
  \centering
  \small
    \begin{tabular}{lcccccccc}
    \textbf{Name} & \textbf{Strands} & \textbf{target} & \textbf{synthesis} & \textbf{decay} & \textbf{physical} & \textbf{PCR} & \textbf{sequencing} & \textbf{read} \\
     & \textbf{synth.} & \textbf{length} & \textbf{method} & \textbf{retained} & \textbf{redundancy} & \textbf{cycles} & \textbf{method} & \textbf{coverage} \\
     \hline
    Goldman & 153335 & 117   & Agilent & 100   & 22172 & 22    & HighSeq 2000 & 519 \\
    Erlich & 72000 & 152   & Twist B. & 100   & 1.28E+7 & 10    & Miseq V4 & 281 \\
    Erlich D1-7 & 72000 & 152   & Twist B. & 100   & 1.28E+7-1.28 & 40+   & Miseq V4 & 281-503 \\
    \origDNA & 4991  & 117   & Customa. & 100   & 3.9E+3 & 65    & Miseq 2x150 & 372 \\
    \origOld & 4991  & 117   & Customa. & 6.25  & 3.9E+4 & 65    & Miseq 2x150 & 456 \\
    \buffer & 4991  & 117   & Customa. & 100   & 1.0   & 68    & Miseq 2x150 & 461 \\
    \bufferOld & 4991  & 117   & Customa. & 5.75  & 17.9  & 68    & Miseq 2x150 & 396 \\
    \end{tabular}%
 \caption{Parameters of the datasets analyzed in this paper.  \label{tab:pardata}}
\end{table}%


\subsection{\label{sec:errinmols}Errors within molecules}

In this section, we analyze the substitution, deletion, and insertion errors in individual molecules. As discussed in Section~\ref{sec:errsources}, those errors are mainly due to synthesis and sequencing. 
We show results for Goldman's and Erlich's dataset, and the \origDNA dataset, as those datasets differ in the synthesis method. Moreover, we show results for our heat treated DNA, \origOld, as decay of DNA introduces substitution errors as well, (as discussed in Section~\ref{sec:storagechemerr}).

\subsubsection{Estimation of overall error probabilities}
We start by considering the aligned reads (obtained from the two single sided reads as explained in Section~\ref{sec:data}), and for those reads, we estimate the substitution, insertion, and deletion probabilities. 
Towards this goal, for each aligned read $\vm$, we find a molecule $\vm_{\mathrm{orig}}$ in the set of original sequences which minimizes the edit (Levenshtein) distance to $\vm$, 
and take the substitution, insertion, and deletion probabilities as the average number of substitutions, insertions, and deletions per nucleotide required for aligning $\vm$ and $\vm_{\mathrm{orig}}$. 
We aggregate the corresponding statistics for all molecules, as well as for molecules that do have the correct target length and do not have the correct target length (Figure~\ref{fig:errorprobs}(a-c)).
Note that molecules that do not have the correct target length necessarily contain substitution or insertion errors. 
Before discussing the error probabilities, we estimate the sequencing error. Note that the overall error consists of errors due to synthesis, storage and sequencing. Out of those error sources, the sequencing error is the only error that can be estimated independently due to two-sided reads. 

\subsubsection{Estimation of reading error probabilities}
In order to get an estimate of the errors due to reading, 
and to understand which proportion of errors in Figure~\ref{fig:errorprobs}(a-c) can be attributed to synthesis and which to sequencing, we perform the following experiment on the data. 
For each dataset considered in this section, we consider the reads that have been successfully aligned, as those are the reads on which the error probability estimates in Figure~\ref{fig:errorprobs})(a-c) are based on, and we want to understand which part of the estimated overall error can be attributed to synthesis and which to sequencing.  
Out of those reads, we consider the two single sided reads and find the best alignment that does not penalize gaps at the beginning and at the end of the sequences. 
We then count the number of substitutions and number of deletions plus number of insertions that are necessary to align the two reads. We then divide that number by the sum of the length of the two single sided reads, and average it over all reads to obtain an estimate of the substitution and insertion plus deletions error probabilities that occur during reading.
This estimate would be the error we obtain when choosing the nucleotide from one of the two single sided reads at random, in case there is a mismatch in the alignment. 
Since we ignore the quality information of the reads, this is a suboptimal way of obtaining a single sided read from the two sided reads, thus the corresponding error probabilities can be viewed as an upper bound on the reading error probability that is achieved by aligning the single sided reads. 
Note that our claim that this is an estimate of the read error probabilities also assumes that the single sided reads are independent. 
The results are depicted in Figure~\ref{fig:errorprobs}(d). 

\subsubsection{Discussion of errors within molecules}

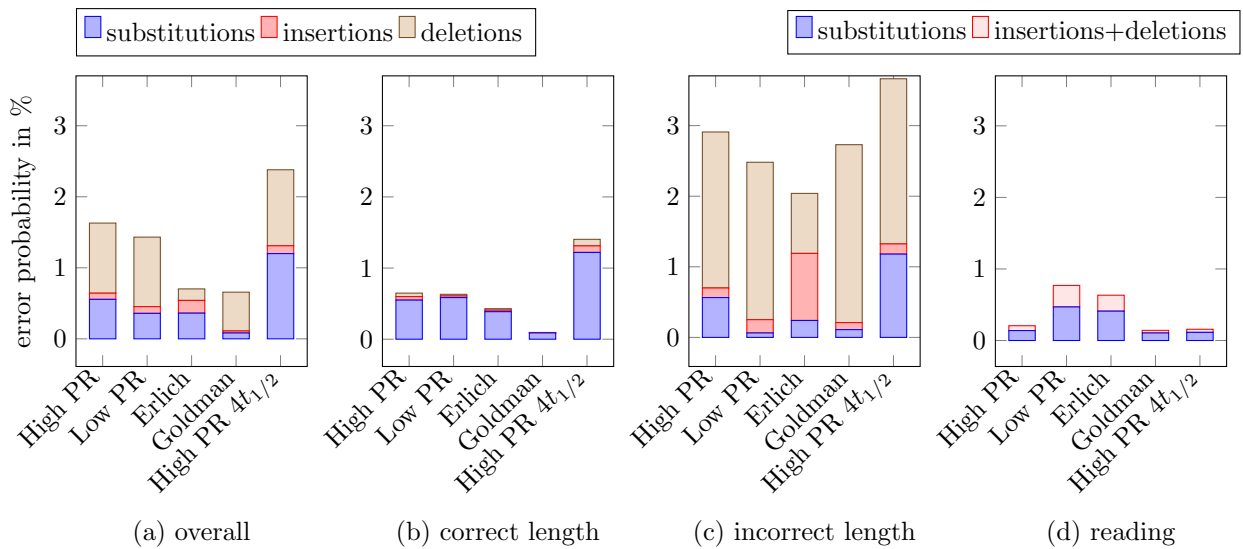
\begin{figure}
\begin{center}
\begin{tikzpicture}
\begin{groupplot}[
title style={at={(0.5,-0.55)},anchor=north}, 
group style={group size=4 by 2, ylabels at=edge left},
         width=0.282\textwidth,
         height=0.33\textwidth,
         ymax=3.23,
]

\nextgroupplot[
	title = {(a) overall},
	ybar stacked,
	bar width=10pt,
    	enlargelimits=0.15,
    	legend style={at={(0,1.15)},
      	anchor=west,legend columns=-1},
    	ylabel={error probability in $\%$},
    	symbolic x coords={1, 2, 3, 4, 5, 6, 7, 8, 9, 10},
    	xtick=data,
    	x tick label style={rotate=45,anchor=east},
    	xticklabels={\origDNA, \buffer, \Erlich,\Goldman, \origOld},
]
       
\addplot+[ybar] plot coordinates {
(1,0.556076)
(2,0.35897399999999996 )
(3,0.36201 )
(4,0.082926 )
(5,1.2)
};    
\addplot+[ybar] plot coordinates {
(1,0.08675800000000002)
(2,0.094445)
(3,0.179013)
(4,0.0281596)
(5,0.11)
};
\addplot+[ybar] plot coordinates {
(1,0.9868740000000001 )
(2,0.978631)
(3,0.162013)
(4,0.5460395999999998)
(5,1.07)
};

\legend{\strut substitutions, \strut insertions, \strut deletions}


\nextgroupplot[
	title = {(b) correct length},
	ybar stacked,
	bar width=10pt,
    	enlargelimits=0.15,
    	legend style={at={(0.5,1.15)},
      	anchor=north,legend columns=-1},
    	symbolic x coords={1, 2, 3, 4, 5, 6, 7, 8, 9, 10},
    	xtick=data,
    	x tick label style={rotate=45,anchor=east},
    	xticklabels={\origDNA, \buffer, \Erlich,\Goldman, \origOld},
]
       
\addplot+[ybar] plot coordinates {
(1,0.55) 
(2,0.585)
(3,0.387) 
(4,0.088)
(5,1.22)
};    
\addplot+[ybar] plot coordinates {
(1,0.049) 
(2,0.023)
(3,0.0211) 
(4,0.0036)
(5,0.092)
};
\addplot+[ybar] plot coordinates {
(1,0.049) 
(2,0.023)
(3,0.0211) 
(4,0.0036)
(5,0.092)
};

\nextgroupplot[
	title = {(c) incorrect length},
	ybar stacked,
	bar width=10pt,
    	enlargelimits=0.15,
    	legend style={at={(0.5,1.15)},
      	anchor=north,legend columns=-1},
    	symbolic x coords={1, 2, 3, 4, 5, 6, 7, 8, 9, 10},
    	xtick=data,
    	x tick label style={rotate=45,anchor=east},
    	xticklabels={\origDNA, \buffer, \Erlich,\Goldman, \origOld},
]

\addplot+[ybar] plot coordinates {
(1,0.564)
(2,0.063)
(3,0.24) 
(4,0.11)
(5,1.18)
};    
\addplot+[ybar] plot coordinates {
(1,0.136) 
(2,0.188) 
(3,0.95) 
(4,0.1)
(5,0.144)
};
\addplot+[ybar] plot coordinates {
(1,2.21)
(2,2.23)
(3,0.85)
(4,2.52)
(5,2.34)
};

\nextgroupplot[
	title = {(d) reading},
	ybar stacked,
	bar width=10pt,
    	enlargelimits=0.15,
    	legend style={at={(-0.9,1.15)},
      	anchor=west,legend columns=-1},
    	symbolic x coords={1, 2, 3, 4, 5, 6, 7, 8, 9, 10},
    	xtick=data,
    	x tick label style={rotate=45,anchor=east},
    	xticklabels={\origDNA, \buffer, \Erlich,\Goldman, \origOld},
]

\addplot+[ybar] plot coordinates {
(1,0.138994)
(2,0.469153)
(3,0.410491)
(4,0.105509)
(5,0.114)
};    
\addplot+[ybar,fill=red!10] plot coordinates {
(1,0.0676675)
(2,0.300715)
(3,0.222388)
(4,0.0363045)
(5,0.043)
};

\legend{substitutions, insertions+deletions}

\end{groupplot}
\end{tikzpicture}
\end{center}
\caption{
\label{fig:errorprobs}
(a) Error probabilities of all molecules, (b) of the molecules that have the correct length (thus the number of insertions is equal to the number of deletions), and (c) error probabilities of molecules that \emph{do not} have the correct length.
In the five datasets, $56.6\%,57.7\%,56.7\%,83\%$, and $78.6\%$ have the correct target length.
All previous works only recovered the information from the molecules that have the correct length. 
In those, the substitution errors dominate by far, which is consistent with previous findings in the literature.
(d) Depicts estimates of the reading errors.
}
\end{figure}

The majority of the reads in each dataset have the correct target length. 
As depicted in Figure~\ref{fig:errorprobs}, out of the reads that have correct target length, in all datasets, the substitution errors dominate by far. 
In the overall error probabilities, the deletions dominate\footnote{At first sight, in Erlich's data, this seems to be different, as the number of insertions is similar to the number of deletions. However, this is only an artifact of how Erlich's reads are filterd: all single-sided reads have length 150, and the target length is 152. Therefore, reads that would align to shorter molecules are filtered out and are not accounted for in the estimates.}. 
Comparing this to the estimated reading error probabilities, we can conclude that likely most of the deletions (and insertions) that we see in the overall reads are due to synthesis, while the substitution errors are likely dominated by synthesis and sequencing and are also impacted by DNA decay and PCR. 
This is consistent with what we expect from the literature, which found that at sequencing, substitution errors are significantly more likely than deletions and insertions (see~Section~\ref{sec:errorsseq}).

\begin{figure}
\begin{center}
\begin{tikzpicture}[scale=1]
\begin{axis}[
  width = 0.8\textwidth,
  height=4.5cm,
  xbar=\empty,
      ylabel=conditional err. prob.,
    enlargelimits=0.05,
    legend style={at={(1.2,1)},
        anchor=north,legend columns=1},
    ybar interval=0.5,
	xticklabels={
	\scriptsize A2C, \scriptsize T2G, 
	\scriptsize A2G, \scriptsize T2C,
	\scriptsize A2T, \scriptsize T2A,
	\scriptsize C2A, \scriptsize G2T,
	\scriptsize C2G, \scriptsize G2C,
	\scriptsize C2T, \scriptsize G2A},
]
\addplot coordinates{ 
(1,0.0606676)
(2,0.0174165)
(3,0.118723)
(4,0.132656)
(5,0.0240929)
(6,0.0214804)
(7,0.0473149)
(8,0.0507983)
(9,0.0223512)
(10,0.030479)
(11,0.268215)
(12,0.205806)
(13,0)
}; 
\addlegendentry{\origDNA}

%
%
\addplot coordinates{
(1,0.0457843)
(2,0.025269)
(3,0.118839)
(4,0.148111)
(5,0.0382787)
(6,0.0287716)
(7,0.052039)
(8,0.0673005)
(9,0.0182637)
(10,0.0165124)
(11,0.25269)
(12,0.188141)
(13,0)
};
\addlegendentry{\buffer}

%

\addplot coordinates{
(1,0.0557851)
(2,0.104339)
(3,0.0764463)
(4,0.0692149)
(5,0.0878099)
(6,0.0635331)
(7,0.0671488)
(8,0.20093)
(9,0.0563017)
(10,0.0490702)
(11,0.0924587)
(12,0.0573347)
(13,0)
};
\addlegendentry{\Erlich}

\addplot coordinates{
(1,0.0188235)
(2,0.0717647)
(3,0.104706)
(4,0.0635294)
(5,0.0505882)
(6,0.0294118)
(7,0.0658824)
(8,0.0270588)
(9,0.105882)
(10,0.0494118)
(11,0.276471)
(12,0.103529)
(13,0)
};
\addlegendentry{\Goldman}

\addplot coordinates{
(1,0.0240829)
(2,0.0106064)
(3,0.0605191)
(4,0.0732468)
(5,0.0162216)
(6,0.0117295)
(7,0.0253307)
(8,0.0269528)
(9,0.0141003)
(10,0.0132269)
(11,0.398053)
(12,0.32593)
(13,0)
};
\addlegendentry{\origOld}

\end{axis}

\end{tikzpicture}
\end{center}
\caption{\label{fig:conderrors}
Conditional error probability for mistaking A for C (A2C) for example.
Note that since this is a conditional error probability (conditioned on an substitution error having occurred), the probabilities bars sum to one. 
}
\end{figure}
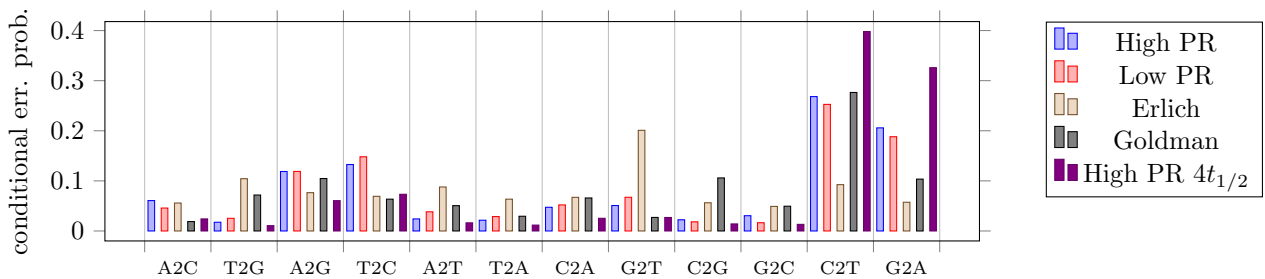

In Figure~\ref{fig:conderrors} we examine the substitution errors in more detail by computing conditional error probabilities for mistaking a certain nucleotide for another. 
The results show that, for example, mistaking T for a C and A for a G is much more likely than other error probabilities. These two transitions can be explained by the vulnerability of cytosine to deamination and the consequent formation of an uracyl. During PCR amplification---either prior to sequencing, or as part of the sequencing process itself---and, if non-proofreading polymerases are used, uracyls are misinterpreted as thymines by the polymerase and adenines are introduced on the complimentary strand (see Figure~\ref{fig:proofreading}). While cytosine deamination is a hydrolysis reaction (involving water as reactant), it can proceed during many of the DNA processing steps, including chemical cleavage of the DNA molecules from the synthesis chip, PCR cycling itself (involving high temperatures), DNA handling and storage. As a result, DNA samples that have undergone decay due to storage show increased C2T and G2A errors. Interestingly, and considering the large experimental differences between the individual datasets (different synthesis, different PCR polymerases, different Illumina sequencing kits, different storage/handling), the observed error levels only display marginal variances, depicting the robustness of the technologies involved. 


\subsection{Distribution of molecules}

In this section we discuss the distribution of the molecules that we observe at the output of the channel. 
The distribution of molecules is a function of the proportions of the molecules in the pool, as well as the sequencing process itself and the number of reads. 
The proportion of the molecules in the pool prior to sequencing depends on the original physical redundancy determined by the synthesis method and potential amplification steps, as well as the decay of the DNA which causes a loss of molecules, and on amplification and potential dilution steps during processing the DNA and for preparation of the sequencing step. 

We are particularly interested in the fraction of sequences observed at least once, since this parameter is relevant for designing coding schemes.
Specifically, as shown in the paper~\cite{heckel_fundamental_2017}, the number of bits that can be stored on a given number of DNA sequences depends on the \emph{fraction of molecules observed at least once}, or in more detail, the fraction out of the molecules that are given to the synthesizer and are read at least once when sequencing the DNA. 

\subsubsection{Distribution of the molecules in control sets and PCR bias}

We start by analyzing Erlich's and Goldman's data as well as our \origDNA and \buffer datasets. 
The main difference in the experiments that generated those datasets is the number of PCR steps as well as the physical redundancy of the molecules.
The synthesis method in some of the experiments differs as well, but the reading coverage and sequencing method in all the experiments are comparable.  
In Figure~\ref{fig:dist_molecules}, we depict the distribution of the number of reads per molecule that has been synthesized, along with the physical redundancy (PR) and the number of PCR cycles (PCRC). 
We find that Erlich's and Goldman's data follows approximately a negative binomial distribution, whereas our datasets (\origDNA and \buffer) have a long tail and peak at zero. 


Since the reading process is essentially the same for all datasets (Illumina sequencing), the difference of Erlich's to Goldman's to our original DNA can in principle be attributed to either a maldistribution of the number of molecules generated during synthesis and/or to PCR amplification bias. 

However, a significant maldistribution of the number of molecules can be ruled by considering Figure~\ref{fig:dist_molecules}(d) and noting that only $20\%$ of the sequences are lost, even though the physical density is around one, and thus the number of DNA fragments in the dataset is approximately equal to the number of distinct fragments that have synthesized. 
In more detail, the corresponding dataset has been obtained by taking a small sample from the original synthesized DNA, which contains many copies of each fragment. 
Suppose the number of copies in the original DNA are very far from being  uniformly distributed. Then, if we draw a small subset corresponding to a physical redundancy of about one, we would expect to lose a significantly larger proportion of fragments. 
As a consequence, we can conclude that the synthesized molecule distribution is not very far from uniform, leaving PCR bias as an possible explanation, as explained next. 

In each PCR cycle, every sequence is amplified with a sequence-specific factor that is slightly below two~\cite{ruijter_amplification_2009,pan_dna_2014,warnecke_detection_1997,caldana_quantitative_2007}. This leads to significantly different proportion of the sequences in the original pool of DNA in particular when the process is repeated many times, i.e., when the number of PCR cycles is large. 
For example, if a sequence has PCR efficiency $80\%$, whereas another one has efficiency $90\%$, then after $60$ PCR cycles, the proportions change from $1/1$ to $(1.8/1.9)^{60}=0.039$. 
Thus, a large number of PCR cycles leads to a distribution with a long tail, see Section~\ref{sec:compexp} for an computational experiment demonstrating this effect. This effect is further confirmed by the dilution experiment from Erlich (Figure 6), where every dilution step consequently requires more PCR cycles until the PCR process saturates.

Finally, note that the dry control set has a significantly larger fraction of lost sequences, see Figure~\ref{fig:dist_molecules}(d), compared to the original DNA in Figure~\ref{fig:dist_molecules}(c). That can be explained by the low  physical redundancy in combination with the PCR amplication bias, as discussed in more detail in the next section.


\definecolor{DarkBlue}{rgb}{0,0,0.7} 
\definecolor{BrickRed}{RGB}{203,65,84}

\begin{figure}[h!]
\begin{center}
\begin{tikzpicture}

\draw[draw,fill=DarkBlue!50,DarkBlue!50] (0,3) -- (14.4,3) -- (0,3.8);
\draw[draw,fill=BrickRed!50,BrickRed!50] (0,4) -- (14.4,4) -- (14.4,3.2);
\node[white] at (2,3.3) {Physical Redundancy};
\node[white] at (13.25,3.6) {PCR cycles};

\begin{groupplot}[
title style={at={(0.5,-0.5)},anchor=north}, 
         group
         style={group size=4 by 3, 
         ylabels at=edge left, yticklabels at=edge left,    
         horizontal sep=0.3cm,vertical sep=2.3cm}, xlabel={\# reads/seq.}, ylabel={frequency},
         width=0.3\textwidth,
]

	\nextgroupplot[title = {\parbox{3.2cm}{\centering (a) \Erlich\\ {\small PR $1.28 10^7$, 10 PCRC} }},] 
	\addplot +[ycomb,DarkBlue!65,mark=none] table[x index=0,y index=1]{./dat/erlich_hist.dat};
	\draw (0,0) circle(1.5pt) -- (1.5cm,1.5cm) node[above right]{$0\%$};

	\nextgroupplot[title = {\parbox{3cm}{\centering (b) \Goldman\\ {\small PR 22172,  22 PCRC} }},] 
	\addplot +[ycomb,DarkBlue!65,mark=none] table[x index=0,y index=1]{./dat/goldmanhist.dat};
	\draw (0,0) circle(1.5pt) -- (1.5cm,1.5cm) node[above right]{$0.0098\%$};

	\nextgroupplot[title = {\parbox{3cm}{\centering (c) \origDNA \\ {\small PR 3898,  65 PCRC} }}, xmax=1500,yticklabels={,,}] 
	\addplot +[ycomb,DarkBlue!65,mark=none] table[x index=0,y index=1]{./dat/S1hist.dat}; 
     	\draw (0,0) circle(1.5pt) -- (1cm,1cm) node[above right]{$1.6\%$};

	\nextgroupplot[title = {\parbox{3cm}{\centering (d) \buffer \\ {\small PR 1.2,  68 PCRC} }},
	,xmax=1500]
            
	\addplot +[ycomb,DarkBlue!65,mark=none] table[x index=0,y index=1]{./dat/GM6hist.dat};
	\draw (0,0) circle(1.5pt) -- (1cm,1cm) node[above right]{$34\%$};

\end{groupplot}          
\end{tikzpicture}
\end{center}

\caption{
\label{fig:dist_molecules}
The distribution of the number of reads per each given sequences that has been synthesized, along with the physical redundancy (PR) and PCR cycles (PCRC). 
Erlich's and Goldman's data is approximately negative binomial distributed, whereas the original DNA and dry control have a long tail and peak at zero. The reading coverage of all datasets is approximately the same. 
Likely, the difference in distribution of (a) and (b) to (c) and (d) is due to the significantly more cycles of PCR in (a) and (b) (see Section~\ref{sec:compexp}), while the difference of (a) and (b) is due to  low physical redundancy and dilution. 
}

\end{figure}
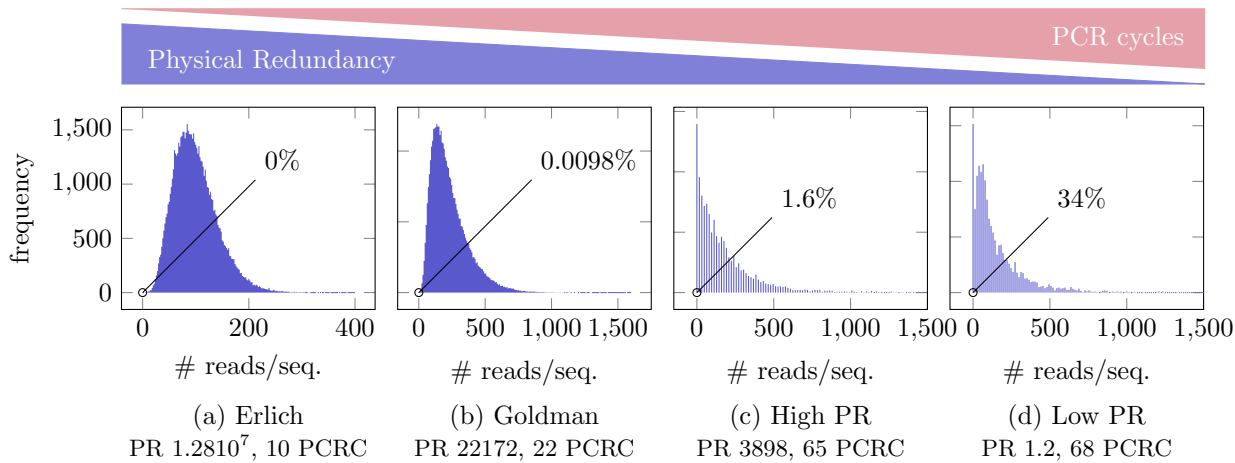


\subsubsection{Low physical redundancy and loss of sequences}

Very low physical redundancy, obtained by taking a subset of a pool of DNA sequences with large physical redundancy, leads to a loss of sequences which in combination with PCR bias (as discussed in the previous section), results in a read distribution that has a long tail and many molecules with few or no reads. This can be seen by comparing Figure~\ref{fig:dist_molecules}(c) with Figure~\ref{fig:dist_molecules}(d). 
To see that the loss of sequences can be partly attributed to generating a pool with low physical redundancy, consider a pool of DNA sequences where each sequence has about the same amount of copies, and that number is large so that the physical redundancy is large. 
Suppose we obtain a pool of DNA with very low physical redundancy (say about one, as in Figure~\ref{fig:dist_molecules}(d)) by taking a random subset of the large pool, via pipetting. 
If the DNA sequences are mixed up well in the pool, that process corresponds to drawing a subset of the pool uniformly at random. Then, the expected fraction of distinct sequences in the subset is about $1-e^{-r}$, where $r$ is the physical redundancy. In other words, at a physical redundancy of $1$, about 36\% of the sequences will be lost.
See again Section~\ref{sec:compexp} for an computational experiment demonstrating this effect.
Moreover, sequencing a pool of DNA with low physical redundancy requires more PCR steps, which, as discussed in the previous section, due to PCR bias, leads to a distribution that has a long tail and further contributes to a loss of sequences.

\begin{figure}[h!]
\begin{center}
\begin{tikzpicture}

\begin{scope}[yshift=-0.7cm]
\draw[draw,fill=DarkBlue!50,DarkBlue!50] (0,3) -- (15.5,3) -- (0,3.8);
\draw[draw,fill=BrickRed!50,BrickRed!50] (0,4) -- (15.5,4) -- (15.5,3.2);
\node[white] at (2,3.3) {Physical Redundancy};
\node[white] at (14.3,3.6) {PCR cycles};
\end{scope}

\pgfplotsset{every tick label/.append style={font=\small}}
\begin{groupplot}[
title style={at={(0.5,-0.6)},anchor=north},
         group
         style={group size=7 by 1, 
         ylabels at=edge left, yticklabels at=edge left,    
         horizontal sep=0.1cm,vertical sep=0.3cm}, xlabel={\# reads/seq.}, ylabel={frequency},
         width=0.225\textwidth,
         height=0.225\textwidth,
]
	\nextgroupplot[title = {\small (a) PR$1.28\, 10^{7}$ }, xmax=2400,yticklabels={,,}] 
	\addplot +[ycomb,DarkBlue!65,mark=none] table[x index=0,y index=1]{./dat/erlich1hist.dat}; 
     	\draw (0,0) circle(1.5pt) -- (0.5cm,1cm) node[above right]{$0.0014\%$};

	\nextgroupplot[title = {\small (b) PR$1.28\, 10^{6}$},xmax=2400]
	\addplot +[ycomb,DarkBlue!65,mark=none] table[x index=0,y index=1]{./dat/erlich2hist.dat}; 
	\draw (0,0) circle(1.5pt) -- (0.5cm,1cm) node[above right]{$0.0014\%$};
	
	\nextgroupplot[title = {\small(c) PR$1.28\, 10^{5}$},xmax=2400] 
	\addplot +[ycomb,DarkBlue!65,mark=none] table[x index=0,y index=1]{./dat/erlich3hist.dat};
	\draw (0,0) circle(1.5pt) -- (0.5cm,1cm) node[above right]{$.0028\%$};

	\nextgroupplot[title = {\small(d) PR$1.28 \, 10^{4}$},xmax=2400] 
	\addplot +[ycomb,DarkBlue!65,mark=none] table[x index=0,y index=1]{./dat/erlich4hist.dat};
	\draw (0,0) circle(1.5pt) -- (0.5cm,1cm) node[above right]{$4.27\%$};
	
	\nextgroupplot[title = {\small(e) PR$1.28 \, 10^{3}$},xmax=2400] 
	\addplot +[ycomb,DarkBlue!65,mark=none] table[x index=0,y index=1]{./dat/erlich5hist.dat};
	\draw (0,0) circle(1.5pt) -- (0.5cm,1cm) node[above right]{$65.1\%$};
	
	\nextgroupplot[title = {\small(f) PR$1.28\, 10^{2}$},xmax=2400] 
	\addplot +[ycomb,DarkBlue!65,mark=none] table[x index=0,y index=1]{./dat/erlich6hist.dat};
	\draw (0,0) circle(1.5pt) -- (0.5cm,1cm) node[above right]{$88.1\%$};
	
	\nextgroupplot[title = {\small(g) PR$1.28$},xmax=2400] 
	\addplot +[ycomb,DarkBlue!65,mark=none] table[x index=0,y index=1]{./dat/erlich7hist.dat};
	\draw (0,0) circle(1.5pt) -- (0.5cm,1cm) node[above right]{$93\%$};
\end{groupplot}          
\end{tikzpicture}
\end{center}

\vspace{-0.4cm}

\caption{
\label{fig:dist_moleculeserlich}
The distribution of the number of reads per each given sequences that has been synthesized for Erlich's dilution experiment, which varies the physical density from approximately $1.28\, 10^7$-$1.28$ copies per molecule synthesized. 
The percentage in the picture corresponds to the percentage of molecules that are not observed in the sequenced data.
The distribution has an increasingly long tail (not depicted), e.g., at physical density $1.28$, there is one sequence that has 98800 reads/sequence.
}

\end{figure}
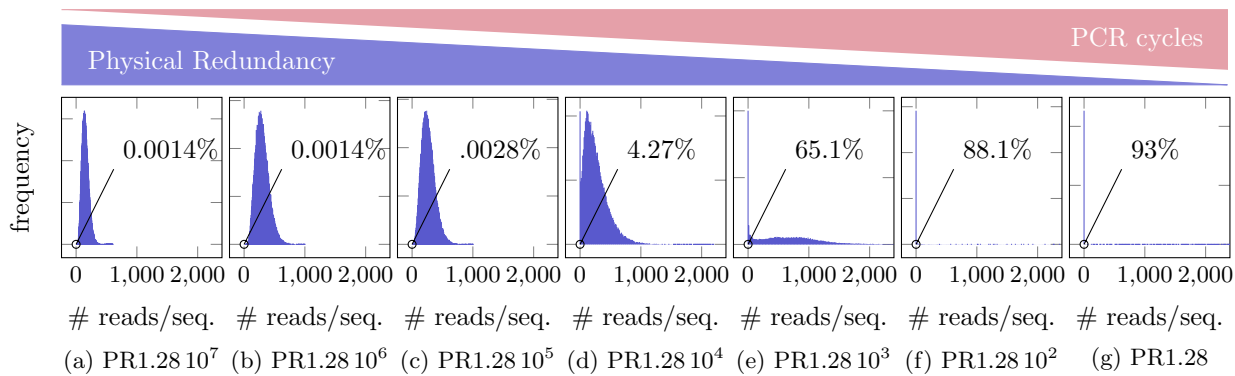

In order to investigate the effect of low physical redundancy further, we next consider the data from an experiment carried out by Erlich and Zielinski~\cite{erlich_dna_2016}, in which the authors varied the physical redundancy from $1.28\, 10^7$-$1.28$ copies per molecule synthesized by factors of 10, by consecutively diluting the DNA.
In Figure~\ref{fig:dist_moleculeserlich}, we depict the corresponding distribution of the molecules. As we can see, initially, when the physical redundancy is large, the molecules are approximately negatively binomial distributed. 
As the physical density becomes low (about 1000 and lower) the fraction of molecules that are not seen becomes increasingly large, and as it becomes very low, a large fraction of the molecules is not seen at the output, and the distribution has an extremely long tail, for example, at physical density $1.28$, there is one sequence that is observed 98800 times at the output. 

Again, this can be attributed to the effect of sampling a small set of sequences from a pool of DNA, as well as to the fact that an increasingly larger number of PCR steps is required for sequencing\footnote{In the Experiment by Erlich and Zielinski~\cite{erlich_dna_2016}, all samples undergo 40 PCR cylces, however, it is to be expected that the PCR reaction saturates approximately 3.3 cycles later for every 10 fold dilution.}, thus the effect of PCR bias becomes more pronounced.

\subsubsection{Errors due to storage}

\begin{figure}[h!]
\begin{center}
\pgfplotsset{every tick label/.append style={font=\small}}
\begin{tikzpicture}[>=latex]

\draw[->] (1.3,2.4)--(1.3,2.7)--(3.8,2.7)--(3.8,2.4);
\node at (2.55,2.9){DNA decay};

\begin{scope}[xshift=5.2cm]
\draw[->] (1.3,2.4)--(1.3,2.7)--(3.8,2.7)--(3.8,2.4);
\node at (2.55,2.9){DNA decay};
\end{scope}


\begin{groupplot}[
title style={at={(0.5,-0.5)},anchor=north}, 
         group
         style={group size=7 by 1, 
         ylabels at=edge left, yticklabels at=edge left,    
         horizontal sep=0.25cm,vertical sep=2.3cm}, xlabel={\# reads/seq.}, ylabel={frequency},
         width=0.24\textwidth,
         height=0.24\textwidth,
]

	\nextgroupplot[title = {\parbox{3cm}{\centering (a) \origDNA\\ {\small PR 3898 
	} }}, xmax=1500,yticklabels={,,},xmax=1400] 
	\addplot +[ycomb,DarkBlue!65,mark=none] table[x index=0,y index=1]{./dat/S1hist.dat}; 
     	\draw (0,0) circle(1.5pt) -- (1cm,1cm) node[above right]{$1.6\%$};
	
	\nextgroupplot[title = {\parbox{3cm}{\centering (b) \origOld \\ {\small PR 3898
	} }},xmax=1400] 
	\addplot +[ycomb,DarkBlue!65,mark=none] table[x index=0,y index=1]{./dat/S3hist.dat}; 	
     	\draw (0,0) circle(1.5pt) -- (1cm,1cm) node[above right]{$8\%$};

%

	
	\nextgroupplot[title = {\parbox{3cm}{\centering (c) \buffer \\ {\small PR 1
	} }},xmax=1400] 
	\addplot +[ycomb,DarkBlue!65,mark=none] table[x index=0,y index=1]{./dat/GM6hist.dat};
	\draw (0,0) circle(1.5pt) -- (1cm,1cm) node[above right]{$34\%$};

	\nextgroupplot[title = {\parbox{3cm}{\centering (d) \bufferOld \\ {\small PR 18
	} }},xmax=1400] 
	\addplot +[ycomb,DarkBlue!65,mark=none] table[x index=0,y index=1]{./dat/GM8hist.dat};
	\draw (0,0) circle(1.5pt) -- (1cm,1cm) node[above right]{$53\%$};
            
\end{groupplot}          
\end{tikzpicture}
\end{center}

\vspace{-0.4cm}

\caption{
\label{fig:dist_molecules_storage}
The distribution of the number of reads per sequence before and after storage. The number of PCR cycles is comparable (65-68 cycles) in all six experiments. In both experiments, the number of lost sequences decreases significantly from the original DNA to the aged one.
}

\end{figure}
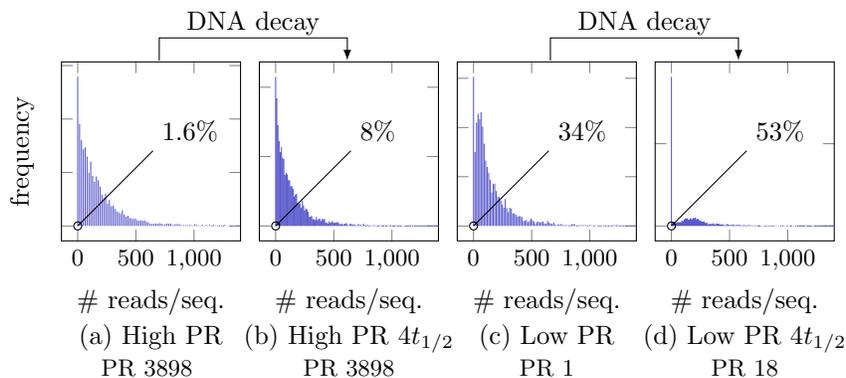

Finally, we examine the effect of storage on the errors by considering the data from our accelerated aging experiments.
See Figure~\ref{fig:dist_molecules_storage} for the corresponding distributions of the molecules.  
We find that, in each experiment, relative to the non-aged data, the number of molecules that are not seen is significantly larger in the aged DNA in the respective dataset compared to the corresponding non-aged dataset.
Specifically, as shown in Figure~\ref{fig:dist_molecules_storage}, 
the number of molecules that are not seen at the output 
in the dataset \origDNA is 1.6\% and increases to 8\% after 4 halftimes (\origOld),
and in the dataset \buffer it is 34\% and increases to 53\% after 4 halftimes (\bufferOld). Note, that the original DNA and the aged DNA were both diluted so that the decayed samples and non-decayed samples required a comparable number of PCR cycles for PCR saturation. 

As the storage related DNA decay process can be best described as a first order reaction, the resulting DNA concentration is expected to decay exponentially~\cite{grass_robust_2015}. 
At one half-life of DNA decay (corresponding to about 500 years for DNA encapsulated in bone~\cite{allentoft_half-life_2012}), half of the DNA has decayed and is no longer amplifiable via PCR. 
To generate DNA with an equivalent thermal history, we have utilized accelerated aging. As non-decayed DNA is diluted prior to amplification and reading to achieve equivalent PCR cycles for the decayed and non-decayed samples, the experiment \buffer versus \bufferOld singles out the effect of aging, since other experimental factors, such as synthesis, amplification and sequencing are kept constant. Figure~\ref{fig:dist_molecules_storage} shows that storage has an effect on the distribution of sequences in the pool, most importantly resulting in an decreased number of sequences that are read at least once. 
Consequently it can be concluded that DNA decay due to storage alters the distribution of the molecules in the pool, which may be attributed to a sequence dependence of the decay process~\cite{pedone_sequence_2009,fujii_sequence_2007,goddard_sequence_2000,hunter_sequence_1993}.

We also found that aging only has a marginal effect on the insertions and deletion probabilities, but increases the substitution error probabilities substantially, as discussed in Section~\ref{sec:errinmols}.

\subsubsection{Impact of PCR bias and processing DNA at low physical redundancy: A computational experiment\label{sec:compexp}}

In this section, we perform a simple experiment demonstrating that PCR bias and processing DNA at low physical redundancies significantly impacts the distribution of the reads. The corresponding results contribute to interpreting and understanding the experiments discussed in the previous sections.

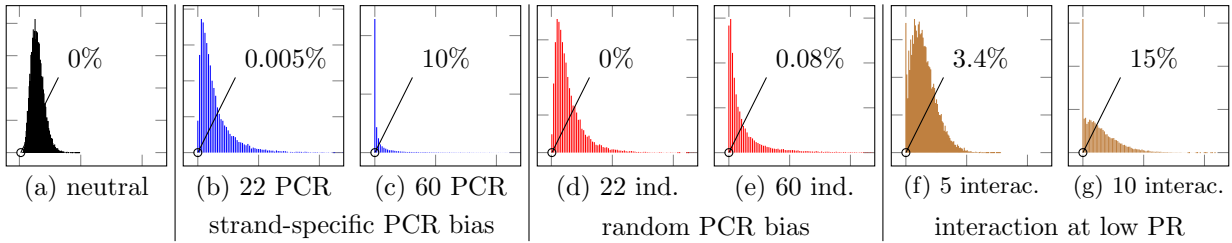
\begin{figure}[h!]
\begin{center}
\begin{tikzpicture}

\draw (2.245,2.15) -- (2.245,-1);
\draw (6.95,2.15) -- (6.95,-1);
\draw (11.655,2.15) -- (11.655,-1);

\node at (4.5975,-0.8){strand-specific PCR bias};
\node at (9.3025,-0.8){random PCR bias};
\node at (14.008,-0.8){interaction at low PR};

\pgfplotsset{every tick label/.append style={font=\small}}
\begin{groupplot}[
title style={at={(0.5,-0.1)},anchor=north}, 
         group
         style={group size=7 by 3, 
         ylabels at=edge left, yticklabels at=edge left,
         xticklabels at=edge bottom,    
         horizontal sep=0.22cm,vertical sep=0.1cm}, 
         width=0.225\textwidth,
         height=0.225\textwidth,
]


	\nextgroupplot[title = {(a) neutral },xmax=2400,yticklabels={,,}] 
	\addplot +[ycomb,black,mark=none] table[x index=0,y index=1]{./dat/dshuffle0.dat};
	\draw (0,0) circle(1.5pt) -- (0.5cm,1cm) node[above right]{$0\%$};
	
	\nextgroupplot[title = {(b) 22 PCR },xmax=2400] 
	\addplot +[ycomb,blue,mark=none] table[x index=0,y index=1]{./dat/dshuffle22.dat};
	\draw (0,0) circle(1.5pt) -- (0.5cm,1cm) node[above right]{$0.005\%$};
	
	\nextgroupplot[title = {(c) 60 PCR },xmax=2400,scaled ticks=false]
	\addplot +[ycomb,blue,mark=none] table[x index=0,y index=1]{./dat/dshuffle60.dat};
	\draw (0,0) circle(1.5pt) -- (0.5cm,1cm) node[above right]{$10\%$};



	\nextgroupplot[title = {(d) 22 ind. },xmax=2400]
	\addplot +[ycomb,red,mark=none] table[x index=0,y index=1]{./dat/shuffle22.dat}; 
	\draw (0,0) circle(1.5pt) -- (0.5cm,1cm) node[above right]{$0\%$};
	
	\nextgroupplot[title = {(e) 60 ind. },xmax=2400]
	\addplot +[ycomb,red,mark=none] table[x index=0,y index=1]{./dat/shuffle60.dat};
	\draw (0,0) circle(1.5pt) -- (0.5cm,1cm) node[above right]{$0.08\%$};

	
	           
	\nextgroupplot[title = {\small (f) 5 interac.},xmax=2400,xticklabels={,,}] 
	\addplot +[ycomb,brown,mark=none] table[x index=0,y index=1]{./dat/dilute4.dat};
	\draw (0,0) circle(1.5pt) -- (0.5cm,1cm) node[above right]{$3.4\%$};              
	\nextgroupplot[title = {\small (g) 10 interac.},xmax=2400,xticklabels={,,}] 
	\addplot +[ycomb,brown,mark=none] table[x index=0,y index=1]{./dat/dilute10.dat};
	\draw (0,0) circle(1.5pt) -- (0.5cm,1cm) node[above right]{$15\%$};  

\end{groupplot}          
\end{tikzpicture}
\end{center}

\caption{
\label{fig:dist_molecul_exp}
The effect of PCR bias and interaction at low physical redundancy on the distribution of the molecules; the y-axis are the number of molecules, and the x axis in each figure is the number of reads per sequence, and varies from 1-2500 in each figure; the longer tails of the distributions have been cut off.
The specific frequency values on the y-axis are not important and are therefore intentionally not show. 
The percentage value in the figures correspond to the percentage of molecules that are not observed in the sequenced data. 
Panel (a) shows the distribution of copies obtained by ``sequencing'' the original pool, and panels (b) and (c) show the distribution after 22 and 60 cycles of PCR, where we took the PCR efficiencies as Gaussian distributed with mean $1.85$ and standard deviation $0.07$.
Panels (d) and (e) depict results from the same experiment, but the efficiencies for each molecule are now different in each cycle and are Gaussian distributed with mean $1.85$ and standard deviation $0.25$. 
Finally, panels (f) and (g) show the distribution after repeated interaction with a pool of small (100) physical redundancy.
The results show that both PCR bias as well as interaction (i.e., repeatedly taking copies of the DNA) at low physical densities leads to a significant loss of fragments.
}
\end{figure}

Towards this goal, we generate a random pool of DNA by taking $\M = 20.000$ sequences, and generate copies of those fragments by drawing from a Gamma distributed with shape parameter $8$ and scale parameter $16$. Thus, the expected physical density in the pool is $8\cdot 16 = 128$.
We artificially sequence from this pool at physical redundancy $300$ by taking $300 \M$ draws with replacement. 
See Figure~\ref{fig:dist_molecul_exp}(a) for the corresponding distribution of molecules.
Note that this implies that the reads are approximately Poisson-Gamma distributed.
Specifically, they are approximately distributed as a mixture of Poisson distributions, with the mean parameter $\gamma$ of the Poisson distribution being Gamma distributed).
A Poisson-Gamma mixture distribution in turn is equivalent to a negative binomial distribution, which has been found to model empirical sequencing distributions well. 
While this justifies our model for generating a random pool of DNA, the particular distribution of the molecules in the pool is not crucial for illustrating how PCR bias impacts the distribution of the molecules, and the conclusions of the computational experiments discussed next are not sensitive to the particular distribution of the molecules in the pool; for example they continue to hold if the molecules in the pool are uniformly distributed.

\paragraph{PCR bias:}
We start by simulating the impact of PCR bias. 
In theory, a single PCR step generates exactly one copy per molecule. In reality, PCR amplifies each fragment on average by a factor that is tightly concentrated around a value slightly smaller than two \cite{ruijter_amplification_2009,pan_dna_2014,warnecke_detection_1997,caldana_quantitative_2007}. In our first experiment, we model the efficiency $E$ of PCR as a Gaussian random variable with mean $1.85$ and standard deviation $0.07$. 
We then amplify each molecule by a factor $E^{N_{\text{PCR}}}$, where $N_{\text{PCR}}$ is the number of PCR cycles. The corresponding distributions, obtained by artificially sequencing as explained above, after $22$ and $60$ such PCR steps, are depicted in Figure~\ref{fig:dist_molecul_exp}(b) and \ref{fig:dist_molecul_exp}(c). 
Note that after a moderate number of PCR steps (22), the distribution is still approximately negative binomial, albeit with a longer tail.
After many PCR steps (60), however, a significant proportion of molecules is not seen and the distribution has a long tail, since the proportions of the molecules in the pool changes drastically, with some frequencies being much more frequent than others. 

Note that this experiment assumes that the PCR efficiency is strand specific and constant at each cycle of PCR. While this has been found to be true in the literature~\cite{ruijter_amplification_2009,pan_dna_2014,warnecke_detection_1997,caldana_quantitative_2007}, we point out that we would observe the same effect even if the PCR efficiency would not be strand specific or would slightly vary at each cycle. 
To see that, we perform the same experiment, but now, at each cycle of PCR, we draw a strand specific DNA efficiency from a Gaussian distribution with mean $1.85$ and standard deviation $0.25$. 
While this is chemically implausible, it shows that the findings from the previous experiment are not sensitive to the assumption of each molecule being duplicated with exactly the same strand-specific factor. 
The corresponding reading distributions, after 22 and 60 such ``molecule independent'' PCR steps is shown in Figure~\ref{fig:dist_molecul_exp}(d) and Figure~\ref{fig:dist_molecul_exp}(e).

\paragraph{DNA storage at low physical redundancy:} 
A common viewpoint is that when storing data on DNA, we can interact with the DNA by reading the DNA many times, or taking copies of the DNA as we please, without significantly changing the pool. 
However, at low physical redundancy, such interactions with the DNA change the distribution of the sequences in the pool significantly, and in particular lead to a significant loss of sequences. 
To demonstrate this effect, we next consider a computational experiment that repeatedly takes a subset of the DNA, and then applies perfect PCR (without a bias) to re-obtain a pool of the original size. This corresponds to making copies of a pool of DNA or reading the pool at low physical redundancy. 
We generate a pool consisting of $100$ copies of $\M = 20.000$ distinct DNA molecules, corresponding to a physical redundancy of $100$.
We then take a small part of the pool and perfectly amplify that part. 
In particular, we choose uniformly at random one tenth of the pool, and then amplifying each sequence by a factor of $10$. 
We then read at coverage $300$ by taking $300\M$ draws with replacement from the pool, as before.
The results after $5$ and $10$ such interaction steps are depicted in Figure~\ref{fig:dist_molecul_exp}(f) and Figure~\ref{fig:dist_molecul_exp}(g).
The results show that repeated interaction with the pool at very low physical densities can lead to a significant loss of the fragments. 
In contrast, interaction at high physical redundancy does have very little effect on the distribution.

\newpage

\section{Conclusion}

In this paper, we characterized the error probabilities of the DNA data storage systems, and assigned, when possible the error sources to processes such as reading and writing the DNA. The key findings are as follows.
\begin{itemize}
    \item Errors within sequences are mainly due to synthesis, sequencing, and to a smaller extent to decay of DNA. Synthesis introduces mainly deletion errors, and potentially a few insertions. If next generation sequencing technology is used for reading, then sequencing introduces additional substitution errors, but only very few insertions or deletions, if at all. 
Long term storage leads to additional substitution errors.     
    Moreover, the majority of the reads has no substitution or deletions errors, thus the errors within sequences are relatively small.
    \item PCR bias significantly changes the distribution of the reads, and a large number of PCR cycles leads to a distribution with a very long tail, which in turn increases the number of  sequences that are never read. 
    \item Storage results in a decay of DNA which results in a significant loss of sequences. In addition, storage increases the substitution error probabilities, due cytosine deamination. 
    \item Storing DNA at low physical redundancy is desirable since it results in a large information density. However, interacting with DNA at low physical redundancy  by copying the pool or sequencing parts of the pool significantly changes the corresponding read distributions and results in a significant loss of sequences. Also, low physical redundancies ask for more PCR cycles, making the system more prone to PCR bias. Hence, if data is stored at low physical redundancies, the anticipated interactions have to be taken into account in the design of the system, in particular the parameters of the error-correcting scheme. 
    Distortion at low physical redundancy may not be observed when working with high physical redundancies, and it is desirable to further understand the sequence dependency of these processes (especially polymerase efficiency and DNA decay) in order to improve the design of DNA storage systems. 
\end{itemize}
The consequences for designing DNA storage systems, in particular error correcting schemes for DNA data storage systems, are as follows. 
An outer code is absolutely crucial in order to account for the loss of entire sequences, which are unavoidable. 
The parameters of the code, i.e., the number of erasures that it can correct should be chosen based on the expected number of lost sequences, which in turn depends on the storage time, physical redundancy, and the interactions anticipated with the DNA (such as PCR steps, number of copies). 
Since the error probability within sequences is generally very low (most reads are error-free), there are three sensible approaches to dealing with them:
i) Errors on a sequence level can be, at least partly, corrected with an inner code.
ii) The reads can algorithmically be combined, in order to obtain an estimate of each sequence. Due to redundant reads some of the errors within sequences can thus be potentially be corrected. 
ii) The outer code can also correct the reads that are erroneous. 
Of course, the three approaches can be combined, for example one can use an inner code to correct some errors, and correct errors that the inner code cannot correct with the outer code, 
or one can only algorithmically combine the fragments, and again correct the remaining errors with the outer code. 
The design choice can be based on the error probabilities that are anticipated from the estimates reported in this paper.


\subsection*{Acknowledgements}
We would like to thank Yaniv Erlich for sharing the sequencing data of their paper~\cite{erlich_dna_2016}.



\printbibliography


\appendix
\section{Detailed description of \buffer and \bufferOld \label{sec:bufferdescription}}
In another dataset from our group, we synthesized again 4991 DNA molecules, each of length 117, containing no homopolymers longer than three. After amplifying the initial library for 30 PCR cycles as described in Grass et al. [Gra+15], we prepared 1 mL samples in Eppendorf tubes containing 70 ng/mL of the DNA in Tris-EDTA buffer (10 mM Tris, 1mM EDTA, pH = 8.0). These high redundancy samples were exposed to 65° C for 20 days to accelerate the decay to $4t_{1/2}$. In contrast to \origDNA and \origOld, we then diluted the original DNA 80,000 times and the DNA after thermal treatment 4,600 times so that the physical redundancies are low, and the concentrations of the two samples are equal (as measured by quantitative PCR). 1 µL of each of the diluted DNA solutions was then taken and amplified for 28 PCR cycles (cycling conditions as above). A gel electrophoresis of the PCR samples was then prepared, and bands at 200 bp were cut out, and purified using QIAquick Gel Extraction Kit. Samples were indexed and prepared for sequencing by a second PCR reaction (10 cycles) using one of the indexed reverse primers each together with the universal forward primer (2FU) followed by purification of individual PCR products by gel electrophoresis as described by Grass et al.~\cite{grass_robust_2015}. We then sequenced the original DNA as well as DNA after thermal treatment corresponding to a decay of four half-lifes (Illumina MiSeq platform (2x150bp) with Truseq sequencing primers(Kit v2). The resulting datasets are denoted by \buffer and \bufferOld.

\end{document}